# New horizons in near-zero refractive index photonics and hyperbolic metamaterials

Michaël Lobet[1,2]*, Nathaniel Kinsey[3], Iñigo Liberal[4], Humeyra Caglayan[5], Paloma A. Huidobro[6,7], Emanuele Galiffi[8], Jorge Ricardo Mejía-Salazar[9], Giovanna Palermo[10,11], Zubin Jacob[12,13], and Nicolò Maccaferri[14,15]*

[1]Department of Physics and Namur Institute of Structured Materials, University of Namur, Rue de Bruxelles 61, 5000 Namur, Belgium

[2]John A. Paulson School of Engineering and Applied Sciences, Harvard University, 9 Oxford Street, Cambridge, MA 02138, USA

[3]Department of Electrical and Computer Engineering, Virginia Commonwealth University, Richmond, VA 23284, USA

[4]Department of Electrical, Electronic and Communications Engineering, Institute of Smart Cities (ISC), Public University of Navarre (UPNA), Pamplona 31006, Spain

[5]Faculty of Engineering and Natural Science, Photonics, Tampere University, 33720 Tampere, Finland

[6]Departamento de Física Teórica de la Materia Condensada and Condensed Matter Physics Center (IFIMAC), Universidad Autónoma de Madrid, E-28049 Madrid, Spain

[7]Instituto de Telecomunicações, Instituto Superior Técnico-University of Lisbon, Avenida Rovisco Pais 1, Lisboa, 1049-001, Portugal

[8]Photonics Initiative, Advanced Science Research Center, City University of New York, New York 10027, USA

[9]National Institute of Telecommunications (Inatel), Santa Rita do Sapucaí 37540-000 MG, Brazil

[10]Department of Physics, NLHT Lab., University of Calabria, 87036 Rende, Italy

[11]CNR NANOTEC-Institute of Nanotechnology, Rende (CS), 87036 Rende, Italy

[12]Elmore Family School of Electrical and Computer Engineering, Purdue University, West Lafayette, IN 47907, USA

[13]Birck Nanotechnology Center, Purdue University, West Lafayette, Indiana 47907, USA

[14]Department of Physics, Umeå University, Linnaeus väg 24, 90187 Umeå, Sweden

[15]Department of Physics and Materials Science, University of Luxembourg, 162a avenue de la Faïencerie, L-1511 Luxembourg, Luxembourg

*michael.lobet@unamur.be; *nicolo.maccaferri@umu.se

**Abstract**

The engineering of the spatial and temporal properties of both the electric permittivity and the refractive index of materials is at the core of photonics. When vanishing to zero, those two variables provide efficient knobs to control light-matter interactions. This perspective aims at providing an overview of the state of the art and the challenges in emerging research areas where the use of near-zero refractive index and hyperbolic metamaterials is pivotal, in particular light and thermal emission, nonlinear optics, sensing applications and time-varying photonics.

**Introduction**

Generating, manipulating, and detecting light are essential actions in photonics that implicitly require interaction with materials. Tracing back to Maxwell's equations, one can identify two physical quantities that are responsible for the interaction of electromagnetic waves with matter: the relative electric permittivity $\varepsilon_r$ acting on the electric properties of matter, and its magnetic counterpart, the relative magnetic permeability $\mu_r$. Both quantities together give the material refractive index $n = \sqrt{\varepsilon\mu}$. Using a wave-light picture, only a few variables are available in the photonics' toolbox. One can either act on the refractive index contrast between materials, as a direct consequence of boundary conditions, or on the time/frequency dispersion of the refractive index. Therefore, over the past years, massive advances in the engineering of $\varepsilon(\vec{r}, t)$, $\mu(\vec{r}, t)$ and $n(\vec{r}, t)$ have been reported in photonics [1–4]. From periodic spatial modulation of the index using photonic crystals [3,5,6] and the simultaneous use of positive and negative permittivity in plasmonics [2], to the nanoscale engineering of the effective index which enabled to reach negative values [7], control over constituent materials has unlocked new regimes of light-matter interactions. Here, we focus on near-zero refractive index (NZI) photonics [8–10] and hyperbolic metamaterials (HMM) [11–17]. The current evolution, as well as new frontiers and future directions and challenges of these two correlated topics are at the core of the current Perspective.

While a new range of fabrication techniques has enabled to generate a negative index, this is in principle possible only over a restricted set of frequencies. As a result, the index undergoes transitions between being positive and negative, opening frequency windows where the index is "near-zero". As suggested by the provided definition of the refractive index in terms of its electric and magnetic constituent, the frequency range where the index has a near-zero response can be retrieved in three different ways (Figure 1a). The refractive index can reach zero by a vanishing electric permittivity, creating the epsilon-near-zero class (ENZ, $\varepsilon \to 0$); by a vanishing magnetic permeability, inducing the mu-near-zero class (MNZ, $\mu \to 0$) or finally by simultaneously vanishing permittivity and permeability, the epsilon-and-mu-near-zero class (EMNZ, $\varepsilon \to 0$ and $\mu \to 0$) [8–10]. These three classes share common properties due to the vanishing index of refraction (Figure 1b), and we can refer to these materials as near-zero-index (NZI) materials. On the one hand, a range of physical quantities tend to infinity, such as the effective wavelength $\lambda$ inside a NZI medium, $\lambda = \frac{\lambda_0}{n} \to \infty$, $\lambda_0$ being the vacuum wavelength, and the phase velocity $v_\varphi = \frac{c}{n}$ with $c$ the speed of light in vacuum. On the other hand, some other quantities tend to zero, such as the wavevector $k$ or the phase difference $\Delta\varphi$ inside the NZI material, leading to a uniform phase distribution. Nevertheless, not all electrodynamical quantities either tend to zero or infinity in a NZI medium. Some quantities depend on the NZI class, i.e., the way one engineers the near-zero index response. For example, the wave impedance $Z = \sqrt{\frac{\mu}{\varepsilon}}$, the group velocity $v_g$ or the related group index $n_g = c/v_g$ present drastically different values according to the NZI class and their specific geometrical implementation [18,19]. The ability to push multiple key parameters to the aforementioned extremes through NZI engineering enabled novel optical phenomena such as perfect transmission through distorted waveguides [18], cloaking [20,21] and inhibited diffraction [22].

When investigating the transition of the relative permittivity around NZI frequency points, a particularly interesting situation led to the definition of hyperbolic metamaterials which can be explained as follow. As briefly mentioned above, plasmonics opened a whole branch of photonics. A surface plasmon polariton (SPP) corresponds to a light-driven collective oscillation of electrons localized at the interface between materials with dielectric ($\varepsilon > 0$) and a metallic ($\varepsilon < 0$) dispersion. If the interface is flat, as in a thin layer, propagating SPP can propagate along the interface. Alternatively, if the interface has a closed shape, such as in a nanoparticle or a nanowire, the coherent electronic vibration is localized, and the excitation is referred to as a localized surface plasmon (LSPs). When multiple metal/dielectric interfaces supporting surface plasmons occur within subwavelength separation, the associated coupled electromagnetic field exhibits a collective, which can be modeled by an effective medium approximation and the dispersion relation presents a unique anisotropic dispersion. More precisely, an effective permittivity tensor $\hat{\varepsilon}$ can be derived such as

$$\hat{\varepsilon} = \begin{pmatrix} \varepsilon_\perp & 0 & 0 \\ 0 & \varepsilon_\perp & 0 \\ 0 & 0 & \varepsilon_\parallel \end{pmatrix}$$

with $\varepsilon_\perp$ ($\varepsilon_\parallel$) the perpendicular (parallel) component with respect to the anisotropy axis, satisfying $\varepsilon_\perp \varepsilon_\parallel < 0$. Consequently, their iso-frequency surface presents a hyperbolic shape (Figure 1c).

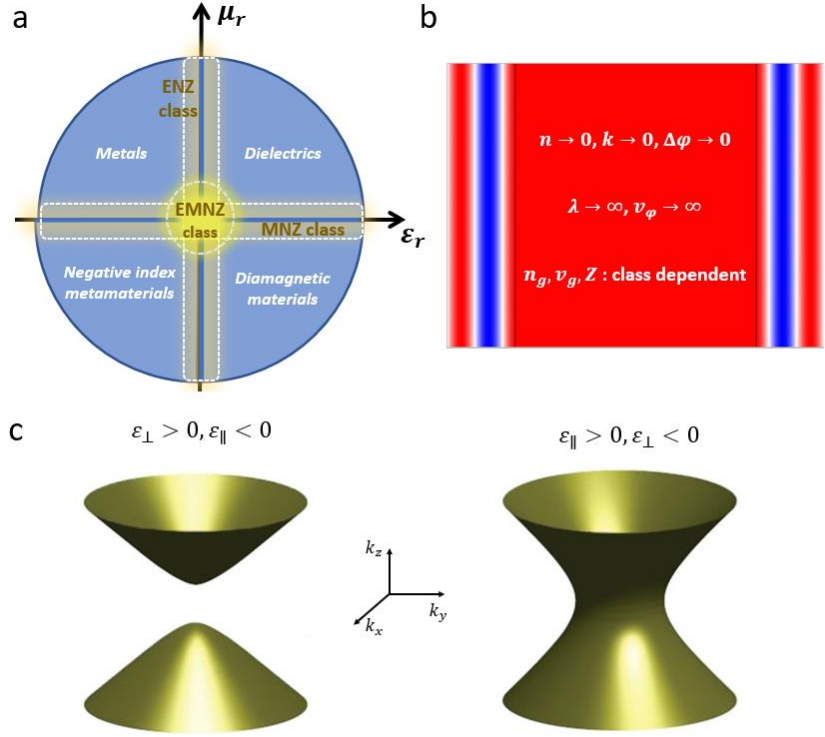

**Figure 1.** (a) Classification of photonic materials according to their relative electric permittivity $\varepsilon_r$ and relative magnetic permeability $\mu_r$, exhibiting three NZI classes: ENZ class, MNZ class and EMNZ class. (b) Uniform phase distribution and electrodynamical quantities reaching extremes values in NZI media. (c) Iso-frequency surfaces in HMMs. Reproduced from Ref. [11].

Those materials, once predominantly engineered artificially, are referred to as hyperbolic metamaterials [11,13,14,17,16]. However, they may occur naturally, too [23–30]. It should be noted that one can engineer the permeability tensor $\hat{\mu}$ in a similar fashion, but this topic will not be covered in the present Perspective, which is structured as follows. We first highlight the impact NZI and HMMs photonics have recently had and are currently having on light and thermal emission. We then move to analyze NZI materials for nonlinear optics and all-optical switching, as well as sensing and magneto-optical applications. We conclude by focusing on the emerging NZI-based time-varying photonics. Overall, our aim is to provide a broad insight into the capabilities and challenges of using these engineered materials to manipulate light-matter interactions in both the frequency and time domain.

**Engineering of light and thermal emission in NZI media**

*Quantum radiative transitions*

NZI media have a profound and nontrivial impact on quantum radiative transitions, e.g., spontaneous emission, stimulated emission, and absorption. Intuitively, one can link the rate of a radiative process with the local density of optical states (LDOS). Then, since a NZI depletes the space of optical modes (Figure 2a), one would be tempted to conclude that NZI media inhibits all radiative transitions, like the band-gap in a photonic crystal. However, this intuitive picture can be misleading. Because the coupling strength also scales with the refractive index, it turns out that a variety of nontrivial radiative phenomena can be observed in the zero-index limit, both as a function of the class of NZI media (ENZ, MNZ, EMNZ) and its effective dimensionality $D$ (3D, 2D, 1D). Specifically, the spontaneous emission decay rate $\varGamma_s$, normalized to its free-space counterpart $\varGamma_0$, scales as follows [19]

$$PF = \frac{\varGamma_s}{\varGamma_0} = Z(\omega)|n^{D-1}(\omega)|.$$

This equation must be evaluated when the transition frequency of the emitter $\omega$ lies in a propagating regime, where both the medium impedance $Z(\omega)$ and the refractive index $n(\omega)$ are real. It illustrates also how a variety of effects can be observed as the refractive index approaches zero (Figure 2b). For example, in three-dimensional media ($D = 3$) the decay rate vanishes independently of the class of NZI media, following the intuition that NZI media depletes the space of optical modes. However, a finite decay rate is obtained for 2D ENZ media and 1D EMNZ media, and the decay rate diverges in 1D ENZ media. The equation above assumes that the emitters are directly coupled to NZI modes, which is accurate only for some metamaterial configurations. Nonetheless, when an emitter is immersed in a continuous medium, one should be careful on accounting for the coupling to the environment, e.g., with the inclusion of local cavity models. The complex interaction of the quantum emitter with surrounding boundaries can lead to further inhibition [31] or enhancement [32] effects. Therefore, very rich emission phenomena arise in NZI media as a function of the class of NZI medium, dimensionality, and how the emitter is coupled to the environment. At the same time, experimental studies of these effects are still rising.

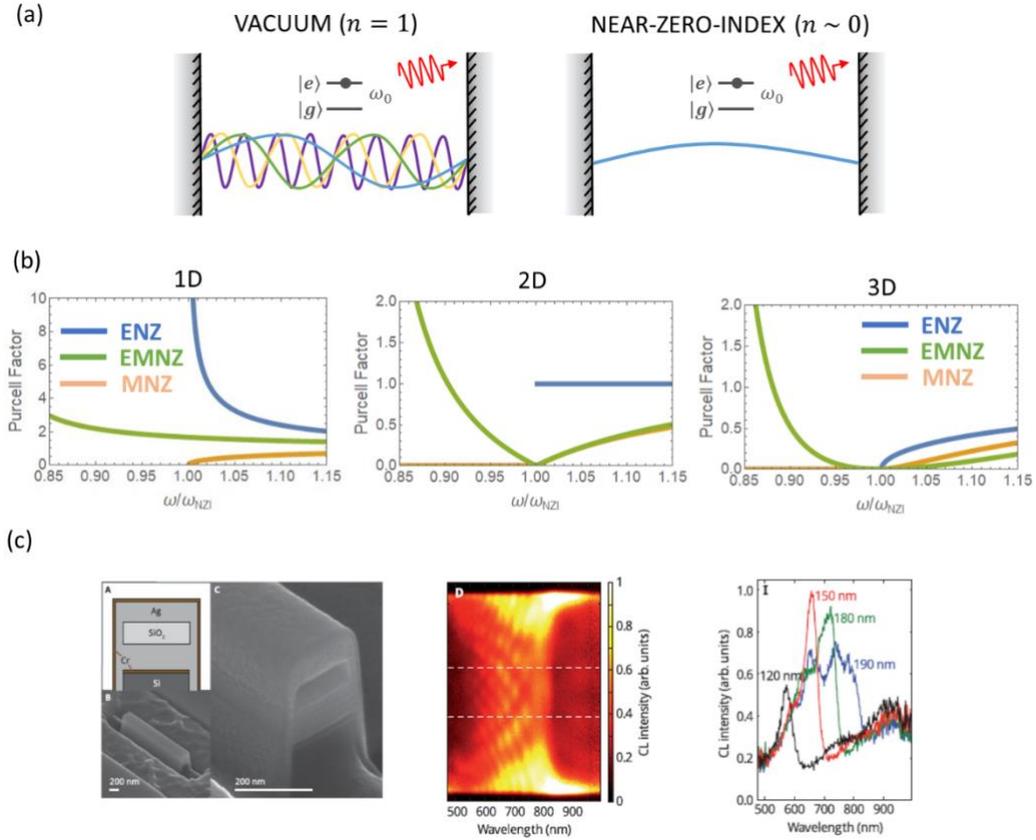

**Figure 2.** (a) Schematic depiction of a two-level system $\{|e\rangle, |g\rangle\}$ with transition frequency $\omega$ coupled to a continuum of photonic modes in a virtual cavity model both in (left) vacuum, and (right) a near-zero-index (NZI) medium that suppresses the spatial density of modes. (b) Purcell factor, $PF = \Gamma_s/\Gamma_0$, in one-dimensional (1D, left), two-dimensional (2D, center) and three-dimensional (3D, right) systems mimicking NZI media with ENZ, MNZ and EMNZ materials properties. Reproduced from Ref. [19]. (c) (Left) SEM image of a rectangular metallic waveguide effectively implementing a 1D ENZ medium at optical frequencies. (Center) Cathodoluminiscence (CL) intensity as a function of wavelength and emission point demonstrating position-independent properties at the effective ENZ wavelength. (Right) CL intensity for different waveguide widths confirming the emission enhancement at the ENZ wavelength. Reproduced from Ref. [34].

1D ENZ media have been experimentally demonstrated at optical frequencies by using metallic rectangular metallic waveguides [33,34]. These experiments have also confirmed both photoluminescence [33] and cathodoluminescence [34] enhancements, exemplifying how 1D ENZ media enhances radiating transitions even in a photonic environment depleted of optical modes. Interestingly, the experiment in [34] also demonstrated position-independent emission, confirming how the enlargement of the wavelength can reduce the accuracy requirements in positioning quantum emitters (Figure 2c).

Engineering spontaneous emission also opens new opportunities for lasing. A photonic crystal laser with parameters compatible with 2D EMNZ media presents a Dirac cone at the Γ point of the Brillouin zone [35]. Their laser is single-mode and remains so as the size of the cavity increases while usually many order modes appear with increasing size. They suggest that the scale-invariant property of the cavity is related to the uniform phase property of NZI environment. The impact of NZI environment on light emission is an interesting avenue for the coming future, especially for designing low-threshold lasers [36] or superradiant lasers [37].

*Applications in quantum technology*

Describing spontaneous emission through a decay rate intrinsically assumes operating in the weak coupling regime and/or under the Markovian approximation [4]. In the weak regime, the emission dynamics follow a simple exponential decay, which can be described by a single parameter, the decay rate and/or lifetime. However, as NZI frequency points typically take place at the edge of a band-gap (or when a band-gap is closed), a wider collection of decay effects can be observed in the nonperturbative regime [38]. In this regime, the decay dynamics can be arbitrarily complex, giving access to a wider range of physical phenomena such as the saturation of the decay rate at a band-edge, the excitation of long-lived bound states, and fractional decay dynamics via the contribution of branch-cut singularities [38]. The importance of these effects, and the interference between them, can be tuned by the design of the shape and size of NZI nanostructures. Interestingly, the possibility of accessing different classes of decay and interaction channels is a convenient tool for quantum simulation, where different physical systems can be implemented and tuned as a function of the dominant nonperturbative decay mechanism [39].

Beyond modifying the individual decay properties of a single emitter, the enlargement of the wavelength in NZI media can trigger collective effects in ensembles of quantum emitters. Thus, NZI media act as optical reservoirs for quantum emitters, which could increase the interaction between optical fields and quantum systems and exhibit enhanced energy transfer and efficient inter-emitter interactions. Several numerical studies have highlighted that NZI media can facilitate the observation of collective effects such as superradiance [40,41], and provide new strategies for entanglement generation [42–46].

Moreover, the concept of entanglement, or non-separability, between qubits is important in various quantum processes such as quantum cryptography and teleportation. While entanglement has traditionally been observed in systems of atoms and ions, it is becoming increasingly accessible in other areas of quantum physics. Specifically, short-distance entanglement has been observed in quantum dots, nanotubes, and molecules, but long-range, i.e., for distance longer compared to the wavelength of light [47,48], qubit-qubit interactions are necessary for long-distance information transfer. In this context, NZI waveguides might represent a gamechanger due to their aforementioned peculiar properties. As examples, numerical studies [42–46] showed that ENZ media outperform the subwavelength distance limitations of qubits cooperative emission in a homogeneous medium. These studies adopted ENZ waveguides into quantum systems, which can be relevant in generating distinctive optical sources, robust entangled states, and other innovative optical applications in different fields of study. It is worth mentioning here that typically electron-phonon, ohmic and inherent losses of the excited ENZ mode, as well as propagation losses, contribute to the transient nature of qubits entanglement mediated by an ENZ medium. Also, the qubit-qubit dissipative coupling induces modified collective decay rates, i.e., superradiant $\Gamma + \Gamma_{12}$ and subradiant states $\Gamma - \Gamma_{12}$, which exhibits pure superradiant emission when $\Gamma = \Gamma_{12}$ condition is satisfied [49]. Here, $\Gamma$ is the decay rate of the individual emitters, while $\Gamma_{12}$ is the modification of the decay rate due to coupling. In summary, the long-range quantum entanglement between a pair of qubits mediated by ENZ waveguide persists over extended periods and long distances. Thus, it is possible to obtain robust entanglement of qubits coupled to the ENZ waveguide channel.

Similar to spontaneous emission, NZI media affects other quantum radiative transitions and light matter interactions. This is particularly exciting for quantum technologies, since achieving strong light-matter coupling in solid-state systems is required for the design of scalable quantum devices. Along this line, it was recently found that dispersion engineering around the ENZ frequency strengthens magnon-photon coupling [50,51]. Strong opto-magnonic coupling would allow for quantum state transfer in hybrid quantum systems. This is a recent and promising direction for NZI materials, and both fundamental and practical implementation advances will be needed to assess the technological potential of NZI media for opto-magnonics.

*Energy vs momentum considerations*

Light-matter interactions are usually described through energetic considerations. However, as noted by Einstein in his seminal work [52,53], momentum deserves an equal theoretical attention due to its conservation property. Examining light-matter interactions inside NZI materials from a momentum perspective [54] therefore offers a different picture. Closely related to the Abraham-Minkowski debate [55–57], light momentum is nontrivial to define. On one hand, Barnett [58] associated Minkowski's momentum to the canonical momentum which is closely correlated to a wavelike nature of light and to the phase refractive index [59]. On the other hand, the Abraham momentum is connected to the kinetic momentum and a particle description of light, represented in equations by the group index. Due to the vanishing index of refraction, NZI induce a vanishing Minkowski momentum. Inhibition of fundamental radiative processes inside 3D NZI can be understood as the impossibility to exchange momentum inside such media [19]. Similarly, diffraction by a slit, which can be seen as a momentum transfer in the direction orthogonal to light propagation is also inhibited [22]. It would be an interesting perspective to generalize those momentum intuitions to other dimensionality of NZI materials [19], especially in the case of the enhanced light-matter interactions in 1D ENZ as described above. Moreover, as pointed out by Kinsey [60], the developed momentum framework could be applied to space-time nonlinear interactions presenting strong spatial and temporal changes. The intriguing regime of these nonlinear responses could benefit from momentum considerations.

*Thermal emission in NZI and HMM media*
Thermal emission is another radiative process of fundamental relevance, which historically was the first to motivate a quantum theory of light. Moreover, thermal emission is also a key process in multiple technologies such as heat and energy management, sensing and communications. However, thermal emission is broadband, temporally incoherent, isotropic and unpolarized, which makes it difficult to control and manipulate. Therefore, different nanophotonic technologies attempt to change these properties by using nanostructured gratings, resonators and/or complex metamaterials [61–63]. Again, because the wavelength is effectively stretched in a NZI medium, it was theoretically demonstrated that the spatial coherence of thermal fields is intrinsically enhanced in NZI media [64]. This interesting result poses a new perspective in engineering thermal emission, where one can enhance the spatial coherence of thermal fields, without the need to resorting to complex nanofabrication processes [64]. In fact, the intrinsic enhancement of thermal emission in ENZ and epsilon-near-pole (ENP) substrates was highlighted by early works in the field of HMM [65]. Hyperbolic media adds a layer of complexity around the ENZ frequency points, resulting in optical topological transitions, where thermal emission can be selectively enhanced or suppressed [66].

Since the medium impedance is enlarged as the permittivity approaches zero, ENZ media naturally acts as high-impedance surface [67] or artificial magnetic conductor [68]. As the tangential electric fields double its strength near a high-impedance surface, ENZ substrates intrinsically enhance the interaction with ultra-thin metallic films. Several prototypes of ultra-thin metallic film thermal emitters have been demonstrated using this principle [69,70]. Moreover, since extreme boundaries are an intrinsic property of NZI media, these emitters have the technological advantage of not requiring from complex nanofabrication processes, and presenting narrowband but spectrally stable emission lines [69,70].

**Nonlinear properties of NZI media and their application to all-optical switching**

Optical switching via nonlinear index modulation has long been a goal of the field, driven by the promise of all-optical devices that are exceptionally fast and operate in environments where electrical control may not be feasible. Through advancements in materials, applications such as saturable mirrors for passive mode-locking [71–73], laser protective eyewear [74,75], and bistable devices [76,77] just to name a few, have been realized, alongside the continual quest to pursue all-optical logic devices [78–80]. For these operations to perform well, devices must effectively modify reflection/transmission/absorption and demonstrate either a latching temporal response or an ultrafast (ideally THz) response, depending upon the use case. In this light, we can turn our attention to the recent developments in ENZ materials and nonlinear optical interactions to consider the advantages and challenges of using ENZ in this sector.

For homogeneous materials, ENZ effects are generally achieved by introducing free carriers, for example, by degenerately doping a semiconductor (e.g. Al:ZnO, In:Sn2O3). In this case, the ENZ condition significantly modifies the dispersion of the material, facilitating strong changes in index even when far from a material resonance (Figure 3a,b) where there may otherwise be minimal dispersion. In this view, ENZ falls into the class of slow-light enhancement schemes for nonlinear optics [81–84] ($n_g$ ~ 2-10 for popular ENZ oxides [85], see Figure 3c), where adding dispersion is used to generate increased light-matter interaction. The nonlinearity in ENZ arises from the modification of the index dispersion either through free-carrier generation (interband effect, blue-shift of index curve) and free-carrier redistribution (intraband effect, red-shift of index curve), see the following for more information [86–89]. In total, ENZ simultaneously improves the absorption of the excitation and provides a steep change in index at a given frequency, which has been shown to facilitate large index modulation on the scale of 0.1 - 1 with ~1 ps relaxation times (Figure 3d-f) [90–92].

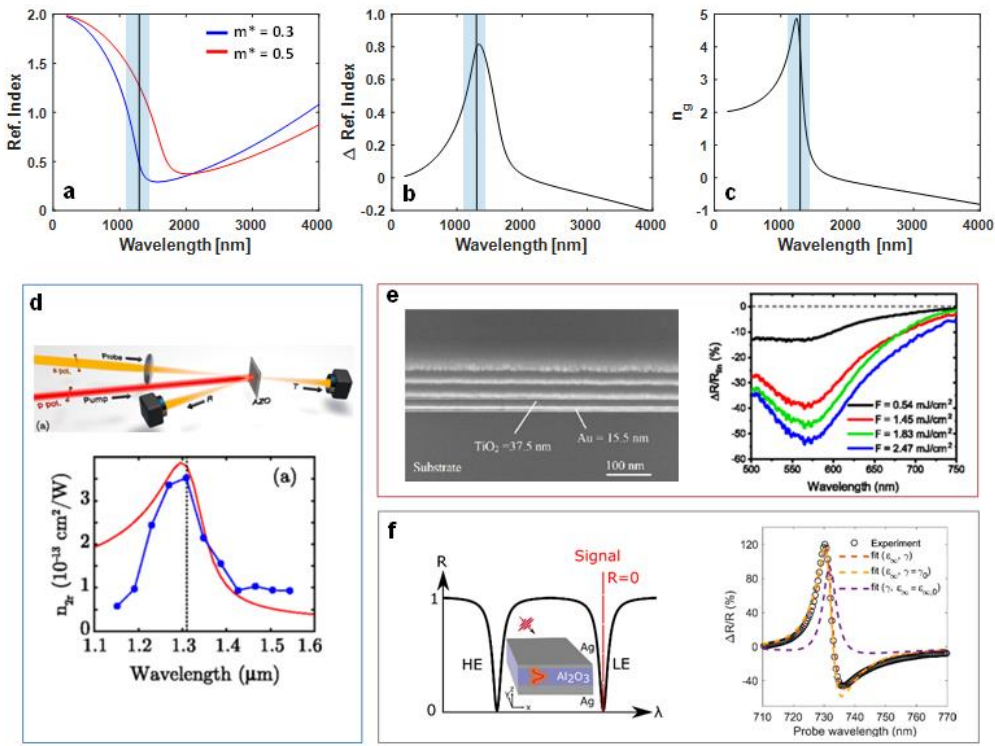

**Figure 3.** (a) Real index of a Drude-based material (blue) with $\varepsilon_\infty = 4, \tau = 6\,fs, N = 8 \times 10^{20} cm^{-3}$ whose effective mass $m^*$ is modulated via intraband nonlinear processes resulting in a shift of the index curve (red), giving rise to a (b) change in refractive index. (c) Group index of the unmodulated Drude-based film as shown in a). The ENZ region is shaded blue with the crossover wavelength indicated as a vertical line. (d) Strong index tuning in Al:ZnO films with ENZ near 1300 nm. Reproduced from Ref. [91] (e) Strong modulation of transmission in effective ENZ materials with crossover at 509 nm. Reproduced from Ref. [93] (f) Modulation of cavity reflection for guided plasmonic mode with mode index near zero. Reproduced from Ref. [94].

To place the performance of ENZ in context, we can compare the nonlinear coefficients to other materials. But before beginning it is important to note that variations in fundamental material and experimental conditions make absolute comparisons a great challenge. As a result, the following is intended to provide a general view on the order of magnitude of responses and trade-offs rather than the specific performance of any given material. Additionally, because nonlinearities in ENZ are non-instantaneous and involve real states (so-called 'slow' processes), they should not be compared to instantaneous nonlinearities involving virtual states (so-called 'fast' processes) as is common, as they are well-known to be much larger [86,95]. A more appropriate comparison is to similar non-instantaneous processes materials such as semiconductors and metals. Finally, while it is common to quantify nonlinearities via $\chi^{(3)}$, $n_2$, or $\alpha_2$, these terms imply properties such as linearity with respect to applied irradiance and an instantaneous response. Such properties are not valid assumptions for the 'slow' nonlinearities in ENZ materials. Thus, we denote the quantities as $\chi^{(3)}{}_{eff}$, $n_{2,eff}$, or $\alpha_{2,eff}$, where subscript 'eff' denotes an *effective* Kerr-like modulation to the optical properties, to highlight that these coefficients do not obey the same rules and depend greatly on properties such as pulse width, applied irradiance, angle of incidence, film thickness, etc.

Table 1 Epsilon-near-zero $n_{2,eff}$ coefficients with associated experimental parameters.

| Material | $n_{2,eff}$ [cm² GW⁻¹]*,† | Relaxation [ps] | Excitation λ [nm] | Probe λ [nm] | Crossover λ [nm] | Pulse Width | Technique |
|---|---|---|---|---|---|---|---|
| Si[101] | $4.5 \times 10^{-5}$ | n/r | 1540 | - | - | 130 fs | Z-scan |
| GaAs [98] | $3 \times 10^{-4}$ | n/r | 1680 | - | - | 111 fs | Z-scan |
| AZO[91] | $3.5 \times 10^{-4}$ | ~0.8 | 785 | 1258 | ~1300 | 100 fs | R/T |
| ITO[102] | $1.80 \times 10^{-3}$ | ~1 | 1100 | 1250 | ~1200 | 150 fs | B.D |
| GZO[96] | $5 \times 10^{-3}$ | ~1 | 1620 | 1700 | 1710 | 60 fs | R/T |
| Au-TiO$_2$[93] | $1.2 \times 10^{-2}$ | ~8 | 470 | 610 | 605 | 120 fs | R/T |
| Ant.-ITO [103] | $-3.7$ | ~1 | 1240 | - | 1240 | 140 fs | Z-scan |

*Note all the values are taken for near normal incidence beams.
† Note that nonlinear index coefficients are functions of the excitation-probe wavelengths, pulse width, sample thickness, irradiance, and angle of incidence. Care should be taken when attempting to use the values outside of the experimental conditions used.
‡ Variations between AZO, GZO, and ITO are largely due to experimental parameter selection (e.g. pump/probe wavelengths) rather than differences in the underlying material.

Now, for ENZ oxides such as Al:ZnO, Ga:ZnO, and In:Sn2O3, $n_{2,eff} = \Delta n/I \sim 0.1 - 5 \times 10^{-3} \, cm^2/GW$ for 1100 - 1700 nm with relaxation on the order of ~1 ps, depending on the wavelength(s) employed [96,97]. This can be compared to free-carrier nonlinearities in the same spectral region for the GaAs platform where $n_{2,eff} \sim 0.1 - 0.3 \times 10^{-3} \, cm^2/GW$ with response times of ~1 ns (crystalline GaAs) [98] that can be reduced to ~ 1 ps for low-temperature grown GaAs [99]. Thus, under optimal excitation conditions, nonlinearities in ENZ oxides provide up to an order of magnitude increase in the strength of the nonlinearity at normal incidence while improving upon the speed. For more information on nonlinear coefficients of various ENZ materials see [100]. It is important to note here that a comparison with virtual processes (for example in semiconductors off-resonance or dielectrics like SiO2) are not appropriate as the mechanisms of the nonlinearity are different and real effects are known to be much larger than their virtual counterparts.

While a useful gain, the introduction of ENZ to modify the dispersion of thin films does not result in a radical performance jump when compared to existing platforms. Additionally, optical loss (due to free carriers) was introduced. As a result, ENZ devices suffer a limited size and must contend with thermal build-up/dissipation that must be addressed to realize high-frequency operation [104–107].

Although the fundamental gains in nonlinearity may not have been extreme, it is important to point out that the primary price paid was loss. In scenarios where devices are small, such loss may not be a large factor in performance (although thermal dissipation remains a concern). As a result, the use of the ENZ region to tailor the dispersion of a material is able to provide an order of magnitude increase in the nonlinearity over competing materials, while maintaining a fast operation, a quite large bandwidth (~400 nm) in the highly relevant telecommunications spectrum, and with readily available materials whose properties can be easily tuned during growth [87]. Additionally, a key benefit of the ENZ oxides is their impressive damage threshold. Routinely, experiments utilize irradiance levels of 10 - 1,000 GW/cm² without permanent damage to the film [88,91,92,100]. This allows ENZ to achieve large absolute changes in the refractive index ($\Delta n \sim 0.1 - 1$), despite only a marginally improved $n_{2,eff}$ value, and consequently, the large absolute changes to reflection, transmission, and absorption at normal incidence that have been observed. With this view, the question becomes, how can we push the strength of the base nonlinearity ($n_{2,eff}$) further to mitigate the need for such high irradiance levels? While gains are predicted when shifting ENZ to the mid-infrared using lower-bandgap materials with lower doping levels [86,108], the tried-and-true method of adding structure is one avenue to continue to engineer the dispersion and improve nonlinear interactions [109–112]. This can be done by structuring the base material (such as forming nano-resonators i.e., meta antennas), coupling the material with a structured layer (such as plasmonic antennas) [113–117], or by mixing multiple materials to achieve an effective ENZ property [93,118–120]. In general, these approaches allow additional freedom to control the dispersion of the device by introducing resonance(s), anisotropy, or both. Recent efforts include coupling to ENZ/Berreman/plasmonic modes within thin layer(s) [118,121–125], incorporating resonant metallic nanoantennas on top of an ENZ layer [103,126,127], and utilizing layered metal-dielectric stacks to produce an effective ENZ condition [93,128]. These techniques can be referred to as HMM and have been successful in reducing the irradiance required to achieve strong control over nonlinear interactions to ~1-10 GW/cm² (a 10-100x reduction), as well as transitioning ENZ into the visible region where natural ENZ materials, such as the doped oxides, are unable to reach. However, these gains are not free. From our view of dispersion engineering, the introduction of structure incurs an additional price of reduced bandwidth (10 - 100 nm), may also require specific excitation conditions (e.g., specific angles of incidence or wavelengths), can lengthen the relaxation time due to nonlinear processes in the added material (e.g. 5-10 ps recovery in metals [129]), and add overall complexity. In total, these undercut some of the key strengths of the ENZ condition, whose ultimate practicality depends upon the constraints of a particular application.

In summary, ENZ condition provides several unique benefits to the nonlinear space founded in the control over material dispersion but also brings baggage in the form of optical loss and only a moderate enhancement. As such, it is not a straightforward solution to the challenges facing nonlinear applications and must be employed appropriately. The primary question facing the community is whether the benefit of ENZ can overcome its limitations and impact an application of relevance. While recent efforts have suggested avenues in pulse characterization [130], frequency shifting [85,126,131,132], bi-stable devices [133,134], and THz generation [135,136], the work is ongoing. We see potential benefits in areas where control over high irradiances is needed or in scenarios where narrow operating bandwidths are utilized, as well as in the use of weakly resonant structures, such as plasmonic antennas, to provide a middle ground wherein the operational spectral bandwidth can remain reasonably broad (~100 nm) while gaining additional improvement to the nonlinearity.

**HMM and ENZ for sensing applications**

The unusual optical properties of HMM have also proven to be useful for optical biosensors with unprecedented levels of sensitivity and resolution [137–139]. Two prototypical HMM, comprising plasmonic nanorod arrays [140,141] and plasmonic/dielectric multilayers [142], are illustrated in Figure 4a,c, respectively. These nanostructures support the so-called volume plasmon polariton (VPP) resonances, which are guided modes resulting from collective excitations of plasmonic resonances in the constituent multilayers [143,144] or nanorods [140,141]. In contrast to conventional surface plasmon polaritons (SPPs), VPPs have their associated electromagnetic fields largely concentrated in the volume of the metamaterial slab and decay exponentially in the superstrate region [140,142,144]. The latter is demonstrated for the nanorod array in the inset of Figure 4a, where simulations of the near-field profile (under VPP resonance) around a single nanorod are shown. This unique feature has inspired two different mechanisms for biosensing applications. First, instead of using continuous flat films, the surfaces of the nanorods can be functionalized with bioreceptors to greatly increase the surface area in contact with the analyte region, producing sensitivity ($S = \Delta\lambda/\Delta n$) values even higher than 40,000 nm/RIU (refractive index unit) [140,141].

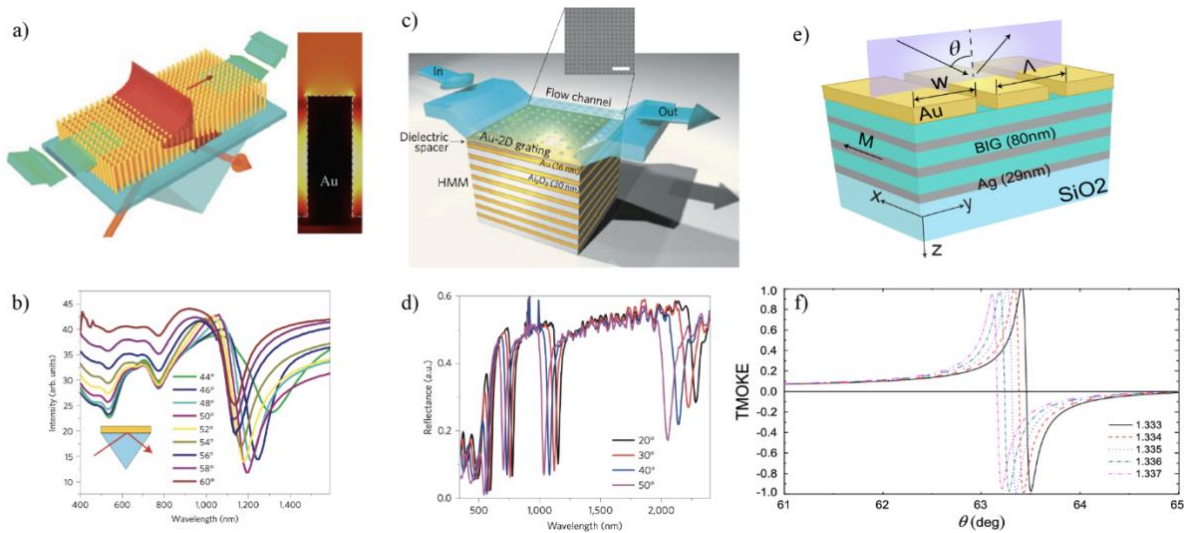

**Figure 4.** (a) Schematic of a conventional Kretschmann-like setup for plasmonic nanorod HMM biosensors and (b) their corresponding reflectance curves for different incident angles. Reproduced from Ref. [140]. The inset in (a), reproduced from Ref. [141], shows the electromagnetic field confinement in the volume of the nanorod array. (c) Illustration of a grating-coupler based multilayer HMM biosensor with a fully integrated fluid flow channel. The inset shows a scanning electron microscopy image of the subwavelength gold diffraction grating on top of the HMM. d) The reflectance spectra for the grating-coupler-HMM at different angles of incidence. Reproduced from Ref. [142]. The blue shift of resonance angles in (b) and (d) with increasing angle of incidence demonstrate that the VPP modes are guided modes. (e) Pictorial view of a MO-HMM comprising dielectric MO layers of bismuth-iron garnet (BIG) and Ag. (f) Fano-like TMOKE curves for the magnetoplasmonic structure in (e) when varying the superstrate refractive index from 1.333 to 1.337. Reproduced from Ref. [145].

Nevertheless, the detection mechanism of plasmonic nanorod metamaterials requires the use of a Kretschmann-like setup, hindering miniaturization due to the need to use bulky prism couplers. Furthermore, plasmonic nanorod metamaterials exhibit a single and relatively broad VPP resonance at the infrared region, as observed from Figure 4b, which also limits the resolution levels. The second biosensing approach considers highly integrable grating-couplers for the excitation of VPPs in plasmonic/dielectric multilayer HMM [142]. Figure 4d shows that various VPP resonances, ranging from infrared to visible wavelengths, are allowed in multilayer HMM. Some of these resonance dips are narrower than the ones for nanorod metamaterials, yielding higher values for the figure-of-merit $FOM = \left(\frac{\Delta\lambda}{\Delta n}\right)\left(\frac{1}{\Delta\omega}\right)$ (where $\Delta\lambda$, $\Delta n$, and $\Delta\omega$ are the resonance shift, refractive index change and full-width of the resonant dip at half-maximum), but with lower sensitivity ($S < 30,000\ nm/RIU$) [142]. A recent proposal combined the advantages of both HMM biosensor configurations into a single structure (by using nanocavities in a multilayer HMM [146], achieving detection limits down to the zeptomole range (i.e., a few tens of molecules).

Despite these breakthroughs, there are still challenges that need to be overcome. For example, the intrinsic ohmic losses of metallic inclusions induce wide resonance curves with large overlaps, which limits resolution when working with ultra-low molecular weight analytes. In addition, biodetection is limited to achiral analytes, making it necessary to use fluorescence-enhanced biosensing techniques for detection of chiral biomolecules [146]. Attempts to surpass these drawbacks include HMMs interfaced with chiral metasurfaces [147], new concepts for manufacturing hyperbolic [113,148,149] and ENZ metamaterials [150], as well as the fabrication of magneto-optical (MO) magnetically-active HMMs [115,151–155]. In MO-HMMs one can take advantage of the transverse MO Kerr effect, with sharp Fano-like curves, to enhance the resolution levels of HMM-based biosensors [145], following a similar approach introduced in the past by Bonanni et al. [156–161]. To illustrate the last mechanism, we consider the grating coupled MO-HMM in Figure 4e, composed by alternating layers of dielectric MO material (BIG in this case) and Ag. Instead of using the reflectance curves (as in conventional non-MO HMM), we may use the TMOKE (as seen from Figure 4f) to reach FOM values as high as 840. In comparison to conventional HMM, achieving FOM up to 590, the use of MO-HMM enables a way to obtain highly enhanced resolution for biosensing applications. Furthermore, computer-aided optimization of the sensor design can be performed with artificial intelligence algorithms, which may not only improve resolution but also the sensitivity of MO-HMM nanostructures [162].

**ENZ media for time-varying photonics**

The possibility of temporally modulating the optical properties of matter via ultrafast optical pumping is establishing a new paradigm for enhanced wave control [163]. While static nanophotonic platforms obey energy conservation and reciprocity, time-modulated systems can overcome these bounds, enabling new functionalities such as nonreciprocity [164–169], frequency generation [170] and translation [171,172], time-diffraction [173], the engineering of photonic gauge fields [174] and synthetic frequency dimensions [175], as well as photonic Floquet matter [176,177], among others. Whilst the field has witnessed dramatic progress at low frequencies, leading to e.g. the first observation of photonic time-reflection [178] and temporal coherent wave control [179], the prospect of unlocking this new wave-control paradigm at near-visible frequencies represents a unique opportunity to broaden and deepen the impact horizon amidst the current rise of photonic technology [180].

Following the pioneering demonstration of the unmatched strength of their nonlinearities [88,92], ENZ media, especially ITO, have gained a spotlight in the quest to implement giant, ultrafast permittivity modulations at near-optical frequencies. Early explorations led to the observation of giant sub-picosecond amplitude modulation via ultrafast shifts of the ENZ frequency of ITO, both by exploiting the coupling to leaky modes [181] and to evanescent ones [121,182] (Figure 5a-b). Currently, efforts are shifting towards using ENZ media as efficient platforms for time-varying wave physics at near-optical frequencies to establish new paradigms for spectral control. Crucially, this endeavour necessarily entails probing the intrinsic modulation speeds available in these materials. A pioneering study demonstrated the temporal analogue of refraction at the interface between two media, a process whereby a change in the refractive index of one of them induces a change in the frequency of light while conserving its momentum [172]. By inducing a large change in the optical properties of a 620 nm ITO film, an extremely broadband and controllable frequency translation of up to 14.9THz was observed in a co-propagating probe (Figure 5c-e). At the quantum level, time-varying ITO in combination with gold nano-antennas has been exploited to spontaneously generate photon pairs from the quantum vacuum [183]. More recently, the temporal analogue of Young's double slit diffraction experiment in photonics was reported [173] (Figure 5f-h), more than fifty years after its prediction [184]. Most remarkably, this experiment revealed the unexpectedly fast nonlinear response of ITO [173], estimating rise times of less than 10 fs, which sparked ongoing theoretical investigations on the nature of such unprecedented response times and the search for new materials exhibiting ultrafast responses of similar timescales. These studies are currently unveiling the key role of momentum conservation in the electron-phonon interaction in such low-electron-density Drude materials, which leads them to support eightfold electron temperatures compared to standard plasmonic materials under analogous illumination conditions (Figure 5i) [185,186].

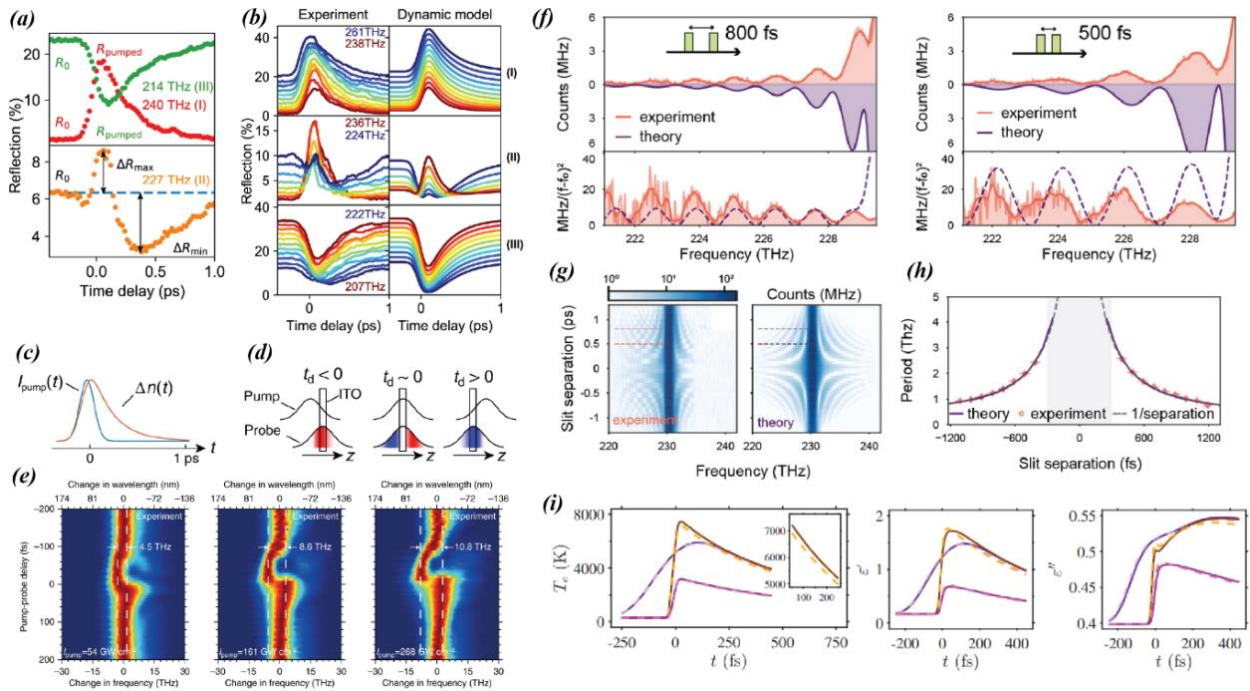

**Figure 5.** (a,b) All-optical switching of an ENZ plasmon resonance in ITO, showing sub-picosecond amplitude modulation of a reflected signal produced by an ultrafast shift in its plasma frequency. Reproduced from Ref. [187] (c-d) Illustration of broadband frequency translation through time refraction in an ENZ material, and (e) its measurement in ITO for increasing pump intensities [172] (f) Experimental measurement (red) and theoretical prediction (blue) of double-slit time diffraction, produced by shining two pump pulses separated by a delay of (left) 800 fs and (right) 500 fs, resulting in accordingly different diffraction fringes. Reproduced from Ref. [181] (g) Experimental (left) and theoretical (right) field intensity from double-slit time diffraction as a function of frequency and slit separation, quantitatively compared in panel (h). (i) Time dependence of (left) the electron temperature, (middle) real and (right) imaginary parts of the ITO permittivity under optical pumping via (purple) a 220-fs pulse at an intensity of 22 GW/cm$^2$, (orange) a 20 fs pulse at 161 GW/cm$^2$ and (magenta) 30 fs at 22 GW/cm$^2$, clearly predicting femtosecond-scale responses in ITO. Reproduced from Ref. [185].

Advances in the quest to achieve single-cycle modulation timescales at near-optical frequencies are further stimulating new theoretical developments towards the efficient modelling of time-varying media. Time-varying effects in subwavelength nanostructures introduce unique challenges [188], as the spatial and temporal scales involved can span several orders of magnitude, and their resolution needs to be comparable in finite-differencing schemes, to ensure numerical stability. In order to overcome adiabatic approximations [172,187], more efficient scattering paradigms and techniques are being steadily developed, including novel approaches to deal with the interplay between temporal dependence and frequency dispersion [189,190]. At the heart of this, however, are fundamental theoretical challenges concerning boundary conditions and conservation laws for electromagnetic fields at temporal inhomogeneities, a field of intense ongoing investigation for basic electromagnetics research [176,191,192].

In turn, these advances in the ultrafast, giant temporal modulation of ENZ media promise a plethora of exciting ideas to be tested in time-varying photonic platforms. Importantly, the possibility of strong modulations at single-cycle timescales may lead to the realization of temporal photonic crystals [193]. Furthermore, other exotic ideas may soon be realized, such as implementing spatiotemporal modulations [194] and non-parametric gain [195,196], chiral pulse amplification [197] or Floquet topological modes [198]. Further possibilities include enhanced emission and mirrorless lasing [193], subdiffractional-mode excitation on non-structured surfaces [199], the spontaneous generation of polariton pairs from the quantum vacuum through the dynamic Casimir effect [200–202], the control over all entanglement degrees of freedom of single photons [203], and the enhancement and tailoring of spontaneous emission of free electrons [204]. Finally, in the context of the topic treated in this section, it is worth closing the circle by making a connection with a topic treated in Section 2. In fact, new opportunities for the engineering of thermal emission are opened when NZI materials are modulated in time [205]. Time-modulation of the refractive index breaks key assumptions in the usual form of the fluctuation dissipation theorem [206] and Kirchhoff's law [207], which form the basis of thermal emitters. Therefore, while thermal fluctuating currents are typically uncorrelated in frequency and space for conventional thermal emitters, time modulation leads to secondary currents that are correlated in frequency and space, opening the door to thermal emission with enhanced coherence and nontrivial photon correlations [208]. Furthermore, energy can be either pumped into a material or retracted from it as it is modulated in time, enabling "active" thermal emitters radiating outside the blackbody spectrum [208], and acting as heat engines [209]. Thermal emission from NZI bodies is particularly sensitive to time modulation. For example, since the near-field of a fluctuating current scale as $E_{NF} \sim 1/(4\pi\varepsilon r^3)$, ENZ bodies support very strong thermal fields within them.

Temporal modulation is capable of releasing these fields, forming the dual of a spatial grating, it consists of a narrowband peak fixed at a given frequency, but whose radiation scans all wave-vectors, from near to far fields [208].

**Conclusions**

We highlighted the tremendous activity of a vibrant research community demonstrating the capabilities of NZI systems and HMM metamaterials to manipulate light-matter interactions in both the frequency and time domain. Engineering of $\varepsilon(\vec{r}, t)$, and consequently $n(\vec{r}, t)$, around their near-zero value broadens the horizons in several areas, including light and thermal emission, nonlinear optics and all-optical switching, as well as sensing and quantum applications. NZI materials are also a promising platform for exploring the emerging field of time-varying photonics. Nevertheless, while providing several unique benefits and demonstrating the above enounced breakthroughs, NZI and HMM research field still face challenges that need to be overcome such as intrinsic ohmic losses of metallic inclusions, reducing its applicability, for instance in sensing. Routes to boost performance of HMM biosensors include the use of nanocavities in multilayer metamaterials (to increase the sensitivity through enhanced electromagnetic field-analyte interactions) or MO effects (to improve resolution). Based on recent developments mentioned in this Perspective, we may foresee the use of plasmonic nanocavities in MO multilayer HMM for future ultrasensitive and ultrahigh resolution biosensors. Moreover, optical forces due to the highly confined electromagnetic fields into deep subwavelength plasmonic nanocavities can provide a way to beat the need to use binding tethers or labelling (e.g., fluorophores) [210–212], improving device recyclability in future developments.

In addition, as we discussed in the last section, ENZ media are also being employed as one of the main platforms for exploring photonics in time-varying media. The underlying reason is their unique capability to provide ultrafast and strong changes of their optical response in the near-IR range through nonlinear effects rooted in nonequilibrium electron dynamics. Thus, ENZ materials provide a ground-breaking platform for exploring new regimes of light-matter interactions. Amidst the quest for translating the growing, rich phenomenology of time-varying media towards the near-visible range, mounting experimental and theoretical evidence points at the prime role that ENZ media will play over the coming years, in turn feeding back new insights into their non-trivial nonequilibrium dynamics.

Finally, ENZ conditions provide several benefits to nonlinear optics thanks to the versatile control over material dispersion. Nevertheless, such a condition implies optical loss and moderate enhancement. We see potential benefits in areas where control over high irradiances is needed or in scenarios where narrow operating bandwidths are utilized, as well as in the use of weakly resonant structures, such as plasmonic antennas, to provide a middle ground wherein the operational spectral bandwidth can remain reasonably broad (~100 nm) while gaining additional improvement to the nonlinearity. To conclude, the fundamental question facing the community is whether the benefit of ENZ condition and hyperbolic dispersion can overcome their limitations to provide relevant applications. Nevertheless, we should look at the future with optimism, as the current advances in the field, in particular in engineering HMM structures for improving sensing capabilities or exploiting ohmic losses in the context of light and thermal emission modulation, as well as recent experimental breakthroughs in the field of time-varying media, make us confident that this field is thriving and will be full of surprises in the upcoming years.

**Authors contribution**

All the authors contributed equally to the writing of the manuscript. M.L. and N.M. led the introduction and conclusions parts, with contributions from I.L and N.K. I.L. led the part on NZI-driven light emission, with contributions from M.L., H.C and Z.J. N.K. led the nonlinear section, with contributions from H.C. and N.M. J. R.M.-S. lead the sensing section with contributions from G.P. and N.M. P.A.H. and E.G. jointly lead the time-varying media section, with contributions from I.L. M.L. and N.M. conceived the project and coordinated the work.

**Acknowledgments**

N.M. acknowledges support from the Swedish Research Council (grant n. 2021-05784), Kempestiftelserna (grant n. JCK-3122), the Wenner-Gren Foundation (grant n. UPD2022-0074) the European Innovation Council (grant n. 101046920 'iSenseDNA'), and European Commission (grant n. 964363 'ProID'). M.L. is funded by the Fund for Scientific Research (F.R.S.-FNRS) of Belgium. N.K. acknowledges support from the National Science Foundation (1808928) and Air Force Office of Scientific Research (FA9550-22-1-0383). I.L. acknowledges support from Ramón y Cajal fellowship RYC2018-024123-I by MCIU/AEI/FEDER/UE, and ERC Starting Grant 948504. H.C. acknowledges the financial support of the European Research Council (Starting

Grant project 'aQUARiUM'; agreement n. 802986). P.A.H. acknowledges support from the Spanish Ministry of Science and Innovation through the Ramón y Cajal program (Grant No. RYC2021-031568-I) and the María de Maeztu Program for Units of Excellence in R&D (CEX2018-000805-M); from the CAM (Y2020/TCS-6545); and from the Fundação para a Ciencia e a Tecnologia and Instituto de Telecomunicações (projects UIDB/50008/2020, 2022.06797.PTDC and UTAP-EXPL/NPN/0022/2021). E.G. acknowledges funding from the Simons Foundation through a Junior Fellowship of the Simons Society of Fellows (855344,EG). J.R.M-S. thanks the financial support from the National Council for Scientific and Technological Development-CNPq (314671/2021-8) and RNP, with resources from MCTIC, Grant No. 01245.010604/2020-14, under the Brazil 6G project of the Radiocommunication Reference Center (Centro de Referência em Radiocomunicações - CRR) of the National Institute of Telecommunications (Instituto Nacional de Telecomunicações - Inatel), Brazil. Z.J acknowledges support from the U.S. Department of Energy (DOE), Office of Basic Sciences (grant n. DE-SC0017717).